\documentclass[10pt]{article}
\usepackage[left=0.95in,right=0.95in,top=0.7in,bottom=0.7in]{geometry}

\usepackage{hyperref}
\usepackage{amsmath, amsthm}
\usepackage{cleveref}
\crefname{ex}{Example}{Examples}
\usepackage[margin=1.5cm]{caption}

\usepackage{tikz, graphicx, pgfplots}
	\pgfplotsset{compat=1.12} 
	\usetikzlibrary{positioning}
	
\usepackage{color, xcolor}
    
	\newcommand\blue[1]{{\textcolor{blue}{#1}}}
	\definecolor{orange}{RGB}{250, 140, 0}
		
	\definecolor{turq}{RGB}{0, 160, 160}
		
	\definecolor{violet}{RGB}{164, 98, 234}

\usepackage{amsthm}
	\newtheorem{thm}{Theorem}[section]

	\theoremstyle{definition}
		\newtheorem{defn}[thm]{Definition}
		\newtheorem{ex}[thm]{Example}
		
	\newtheoremstyle{TheoremNum}
        {\topsep}{\topsep}              
        {\itshape}                      
        {}                              
        {\bfseries}                     
        {.}                             
        { }                             
        {\thmname{#1}\thmnote{ \bfseries #3}}
    \theoremstyle{TheoremNum}

\newcommand{\df}[1]{{\bf\emph{#1}}}		

\usepackage[normalem]{ulem}	

\usepackage{parskip}
    \makeatletter
    \def\thm@space@setup{%
        \thm@preskip=\parskip \thm@postskip=0pt
    }\makeatother
\usepackage{amsfonts, amssymb, amsrefs, mathtools}
\newcommand{\eq}[1]{\begin{align*}#1\end{align*}}
	
\newcommand\mrm[1]{\mathrm{#1}}
 
\newcommand{\rr}{\ensuremath{\mathbb{R}}}

\renewcommand{\epsilon}{\varepsilon}	
\renewcommand{\phi}{\varphi}			

\newcommand{\kk}{\kappa}

\newcommand{\vv}[1]{{\boldsymbol{#1}}}  

\newcommand{\rrpp}{\rr_{+}}

\newcommand{\xx}{\vv x}
\newcommand{\yy}{\vv y}

\usepackage{enumitem}

\usepackage[version=4]{mhchem}
\usepackage{chemfig}

\usepackage[normalem]{ulem}

\usepackage{authblk}
\title{
    Weakly reversible mass-action systems\\ with infinitely many positive steady states
}
\author[1]{
        Bal\'azs Boros%
}
\author[1,2]{
         Gheorghe Craciun%
}
\author[1]{
        Polly Y.\ Yu%
}
\affil[1]{\small Department of Mathematics, University of Wisconsin-Madison}
\affil[2]{\small Department of Biomolecular Chemistry, University of Wisconsin-Madison}
\date{} 

\begin{document}
\maketitle

\begin{abstract}
    We show that weakly reversible mass-action systems can have a continuum of positive steady states, coming from the zeroes of a multivariate polynomial. Moreover, the same is true of systems whose underlying reaction network is reversible and has a single connected component. In our construction, we relate operations on the reaction network to the multivariate polynomial occurring as a common factor in the system of differential equations. 
\end{abstract}

\section{Introduction}
\label{sec:intro}

It was widely believed that every weakly reversible mass-action system has a nonzero, finite number of positive steady states in each invariant polyhedron. One of the fundamental questions about dynamical systems is the description of the set of steady states. Many mathematical analyses implicitly assume that steady states form a discrete set. The authors of \cite{deng:jones:feinberg:nachman:2011} proposed a proof of both existence and finiteness of positive steady states of weakly reversible mass-action systems. Their proof concerning existence was made complete in \cite{boros:2019}; however, their argument on finiteness is insufficient, and as shown in this paper, is not true in general. We construct weakly reversible (even reversible) mass-action systems with continua of positive steady states within some invariant polyhedra.

Some mass-action systems are known to have at most one positive steady state within each invariant polyhedron. For example, detailed-balanced and complex-balanced systems~\cite{horn:jackson:1972}, injective systems~\cite{craciun:feinberg:2005}, and those satisfying the hypotheses in the Deficiency-One Theorem~\cite{feinberg:1995}.

For a weakly reversible system, we expect the set of positive steady states in each invariant polyhedron to be bounded. This follows from the \emph{Permanence Conjecture}~\cite{craciun:nazarov:pantea:2013}: every invariant polyhedron of a weakly reversible mass-action system admits a compact global attractor. The conjecture is proven for some special cases, including in $2$-dimension~\cites{craciun:nazarov:pantea:2013,pantea:2012}, or when the network has a single connected component~\cites{gopalkrishnan:miller:shiu:2014, anderson:cappelletti:kim:nguyen:2018, boros:hofbauer:2019}. Furthermore, since any system of polynomial equations has finitely many nondegenerate positive roots~\cites{khovanskii:1980, khovanskii:1991, bihan:sottile:2007, sottile:2011}, necessarily, the positive steady states in our constructions lie in a bounded set (of an invariant polyhedron), and infinitely many of them are degenerate.

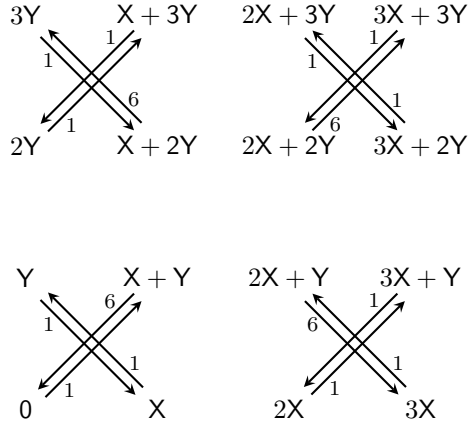
\begin{figure}[h!t!]
    \centering
    \begin{tikzpicture}[scale=1.75]
    \begin{scope}
\node (00) at (0,0) {$\sf{0}$};
\node (10) at (1,0) {$\sf{X}$};
\node (11) at (1,1) {$\sf{X}+\sf{Y}$};
\node (01) at (0,1) {$\sf{Y}$};
\draw[arrows={-stealth},thick,transform canvas={xshift=0.5mm, yshift=-0.5mm}] (00) to node [near start,below] {\footnotesize  $1$} (11);
\draw[arrows={-stealth},thick,transform canvas={xshift=-0.5mm, yshift=0.5mm}] (11) to node [near start,above] {\footnotesize  $6$} (00);
\draw[arrows={-stealth},thick,transform canvas={xshift=0.5mm, yshift=0.5mm}] (10) to node [near start,right] {\footnotesize  $1$} (01);
\draw[arrows={-stealth},thick,transform canvas={xshift=-0.5mm, yshift=-0.5mm}] (01) to node [near start,left]  {\footnotesize  $1$} (10);

\node (20) at (2,0) {$2\sf{X}$};
\node (30) at (3,0) {$3\sf{X}$};
\node (31) at (3,1) {$3\sf{X}+\sf{Y}$};
\node (21) at (2,1) {$2\sf{X}+\sf{Y}$};
\draw[arrows={-stealth},thick,transform canvas={xshift=0.5mm, yshift=-0.5mm}] (20) to node [near start,below] {\footnotesize  $1$} (31);
\draw[arrows={-stealth},thick,transform canvas={xshift=-0.5mm, yshift=0.5mm}] (31) to node [near start,above] {\footnotesize  $1$} (20);
\draw[arrows={-stealth},thick,transform canvas={xshift=0.5mm, yshift=0.5mm}] (30) to node [near start,right] {\footnotesize  $1$} (21);
\draw[arrows={-stealth},thick,transform canvas={xshift=-0.5mm, yshift=-0.5mm}] (21) to node [near start,left]  {\footnotesize  $6$} (30);

\end{scope}

\node (02) at (0,2) {$2\sf{Y}$};
\node (12) at (1,2) {$\sf{X}+2\sf{Y}$};
\node (13) at (1,3) {$\sf{X}+3\sf{Y}$};
\node (03) at (0,3) {$3\sf{Y}$};
\draw[arrows={-stealth},thick,transform canvas={xshift=0.5mm, yshift=-0.5mm}] (02) to node [near start,below] {\footnotesize  $1$} (13);
\draw[arrows={-stealth},thick,transform canvas={xshift=-0.5mm, yshift=0.5mm}] (13) to node [near start,above] {\footnotesize  $1$} (02);
\draw[arrows={-stealth},thick,transform canvas={xshift=0.5mm, yshift=0.5mm}] (12) to node [near start,right] {\footnotesize  $6$} (03);
\draw[arrows={-stealth},thick,transform canvas={xshift=-0.5mm, yshift=-0.5mm}] (03) to node [near start,left]  {\footnotesize  $1$} (12);

\node (22) at (2,2) {$2\sf{X}+2\sf{Y}$};
\node (32) at (3,2) {$3\sf{X}+2\sf{Y}$};
\node (33) at (3,3) {$3\sf{X}+3\sf{Y}$};
\node (23) at (2,3) {$2\sf{X}+3\sf{Y}$};
\draw[arrows={-stealth},thick,transform canvas={xshift=0.5mm, yshift=-0.5mm}] (22) to node [near start,below] {\footnotesize  $6$} (33);
\draw[arrows={-stealth},thick,transform canvas={xshift=-0.5mm, yshift=0.5mm}] (33) to node [near start,above] {\footnotesize  $1$} (22);
\draw[arrows={-stealth},thick,transform canvas={xshift=0.5mm, yshift=0.5mm}] (32) to node [near start,right] {\footnotesize  $1$} (23);
\draw[arrows={-stealth},thick,transform canvas={xshift=-0.5mm, yshift=-0.5mm}] (23) to node [near start,left]  {\footnotesize  $1$} (32);

    \end{tikzpicture}
    \caption{A reversible mass-action system with infinitely many positive steady states. See \Cref{ex:cross} for more details.}
    \label{fig:diag-rev}
\end{figure}

For example, the reversible mass-action system in \Cref{fig:diag-rev} admits infinitely many positive steady states, including all positive zeroes of the common factor in its system of differential equations 
\eq{ 
        \dot{x} &= (x^2y^2+x^2+y^2+1-5xy)[1-xy+y-x], \\
        \dot{y} &= (x^2y^2+x^2+y^2+1-5xy)[1-xy-y+x].
    }
The four positive monomials in the common factor correspond to four translated copies of a base unit, while the negative monomial corresponds to a flipped copy of that (\Cref{fig:diag-rev}). More details of these operations on networks are given in \Cref{sec:operations}. The examples presented in \Cref{sec:examples} are $2$-dimensional, but the construction works in higher dimensions as well.

The rest of this paper is organized as follows. After a short section on mass-action systems (\Cref{sec:massaction}), we present the basic operations used in our construction (\Cref{sec:operations}). Then we construct a number of examples with infinitely many positive steady states (\Cref{sec:examples}). Finally, we list a couple of open questions and conclude with some remarks (\Cref{sec:open}).

\section{Mass-action systems}
\label{sec:massaction}

In this section we briefly introduce mass-action systems and related notions that are necessary for our exposition. For more details about mass-action systems, see \cites{feinberg:1987, gunawardena:2003, angeli:2009, craciun:yu:2018}. The symbol $\rrpp$ denotes the set of positive real numbers.

\begin{defn}
\label{def:crn}
	A {\df{Euclidean embedded graph}} (or a \df{reaction network}) is a directed graph $(V,E)$, where $V$ is a finite subset of $\rr^n$.
\end{defn}

Vertices are points in $\rr^n$, so an edge $(\yy_i, \yy_j) \in E$ can be regarded as a bona fide vector in $\rr^n$. We associate a {\df{reaction vector}} $\yy_j - \yy_i \in \rr^n$ to each edge $(\yy_i, \yy_j)$.

\begin{defn}
\label{def:mas}
	A {\df{mass-action system}} is a triple $(V,E, \kk)$, where $(V,E)$ is a reaction network and $\kk \in \rrpp^{E}$ is a vector of \df{rate constants}. Its \df{associated dynamical system} on $\rrpp^n$ is
    \eq{
        \dot{\xx}
        \,\, = \sum_{(\yy_i,\yy_j) \in E} \kk_{ij} \xx^{\yy_i} (\yy_j - \yy_i),
    }
    where $\xx^\yy = x_1^{y_1}x_2^{y_2}\cdots x_n^{y_n}$. 
    The \df{set of positive steady states} of the mass-action system is 
    \eq{ 
            \left\{\xx \in \rrpp^n \colon  \sum_{(\yy_i,\yy_j) \in E} \kk_{ij} \xx^{\yy_i} (\yy_j - \yy_i) = \vv 0  \right\}. 
    }
\end{defn}
\bigskip 

Since $\dot{\xx}$ lies in the \df{stoichiometric subspace}
    \eq{ 
        S = \mrm{span} \{ \yy_j - \yy_i \colon (\yy_i, \yy_j) \in E\},
    }
the solution of the dynamical system with initial value $\xx_0\in\rrpp^n$ is confined to the invariant polyhedron $(\xx_0+S)\cap \rrpp^n$. Consequently, when we speak about the number of positive steady states of the mass-action system, we typically refer to those within one of these translates.  In all our examples, $S = \rr^n$, and there is only one invariant polyhedron, namely $\rrpp^n$. 

Two network properties are of special interest for our purpose.

\begin{defn}
\label{def:rev}
	A mass-action system $(V,E,\kk)$ is 
	\begin{enumerate}[label=(\alph*)]
	    \item \df{reversible} whenever $(\yy_i,\yy_j) \in E$ if and only if  $(\yy_j, \yy_i) \in E$, or 
	    \item \df{weakly reversible} whenever every edge is part of a directed cycle.
	\end{enumerate}
\end{defn}

We conclude this section by illustrating the notions introduced.

\begin{ex}
\label{ex:MAS}
    Consider the mass-action system
    \begin{center}
    \begin{tikzpicture}[scale=1.75]
    
\node (00) at (0,0) {$\sf{0}$};
\node (10) at (1,0) {$\sf{X}$};
\node (11) at (1,1) {$\sf{X}+\sf{Y}$};
\node (01) at (0,1) {$\sf{Y}$};

\draw[arrows={-stealth},thick] (00) to node [midway, below] {$\kk_1$} (10);
\draw[arrows={-stealth},thick] (10) to node [midway, right] {$\kk_2$} (11);
\draw[arrows={-stealth},thick] (11) to node [midway, above] {$\kk_3$} (01);
\draw[arrows={-stealth},thick] (01) to node [midway, left] {$\kk_4$} (00);


    \end{tikzpicture}
    \end{center}
whose Euclidean embedded graph (or reaction network) is 
\begin{center}
    \begin{tikzpicture}[scale=0.94]
    
\draw [step=1, gray, very thin] (0,0) grid (3.5,3.5);
\draw [ ->, black!70!white] (0,0)--(3.5,0);
\draw [ ->, black!70!white] (0,0)--(0,3.5);

\begin{scope}[shift={(0,0)}]

\node[inner sep=0,outer sep=1] (A) at (0,0)  {\footnotesize \blue{$\bullet$}};
\node[inner sep=0,outer sep=1] (B) at (1,0)  {\footnotesize \blue{$\bullet$}};
\node[inner sep=0,outer sep=1] (C) at (1,1) {\footnotesize \blue{$\bullet$}};
\node[inner sep=0,outer sep=1] (D) at (0,1) {\footnotesize \blue{$\bullet$}};

\draw[arrows={-stealth},very thick,blue,below] (A) to node {} (B);
\draw[arrows={-stealth},very thick,blue,right] (B) to node {} (C);
\draw[arrows={-stealth},very thick,blue,above] (C) to node {} (D);
\draw[arrows={-stealth},very thick,blue,left]  (D) to node {} (A);

\end{scope}


    \end{tikzpicture}
\end{center}
and whose associated dynamical system on $\rrpp^2$ is
    \eq{ 
        \dot{x} &= \kappa_1-\kappa_3 xy, \\
        \dot{y} &= \kappa_2 x-\kappa_4 y.
    }
This weakly reversible (but not reversible) mass-action system has a unique positive steady state. 
\end{ex}

\section{Three operations on mass-action systems}
\label{sec:operations}

In this section, we introduce via examples three operations on mass-action systems, namely, \emph{translation}, \emph{scalar multiplication}, and \emph{addition}. In \Cref{sec:examples}, we will combine these to construct examples of mass-action systems with infinitely many positive steady states.

\subsection{Translation}

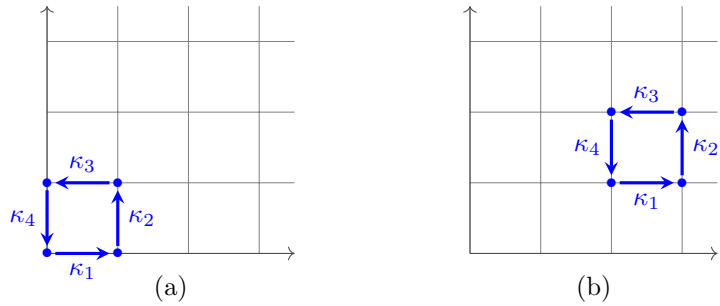
\begin{figure}[ht]
    \centering
    \begin{tikzpicture}[scale=0.94]
    \node at (1.75,-0.5) {(a)};

\draw [step=1, gray, very thin] (0,0) grid (3.5,3.5);
\draw [ ->, black!70!white] (0,0)--(3.5,0);
\draw [ ->, black!70!white] (0,0)--(0,3.5);

\begin{scope}[shift={(0,0)}]

\node[inner sep=0,outer sep=1] (A) at (0,0)  {\footnotesize \blue{$\bullet$}};
\node[inner sep=0,outer sep=1] (B) at (1,0)  {\footnotesize \blue{$\bullet$}};
\node[inner sep=0,outer sep=1] (C) at (1,1) {\footnotesize \blue{$\bullet$}};
\node[inner sep=0,outer sep=1] (D) at (0,1) {\footnotesize \blue{$\bullet$}};

\draw[arrows={-stealth},very thick,blue,below] (A) to node {$\kappa_1$} (B);
\draw[arrows={-stealth},very thick,blue,right] (B) to node {$\kappa_2$} (C);
\draw[arrows={-stealth},very thick,blue,above] (C) to node {$\kappa_3$} (D);
\draw[arrows={-stealth},very thick,blue,left]  (D) to node {$\kappa_4$} (A);
\end{scope}

    \end{tikzpicture}
        \hspace{2cm}
    \begin{tikzpicture}[scale=0.94]
    \node at (1.75,-0.5) {(b)};

\draw[arrows={-stealth},very thick,white,below] (0,0) to node {\vphantom{$\kappa_1$}} (1,0);

\draw [step=1, gray, very thin] (0,0) grid (3.5,3.5);
\draw [ ->, black!70!white] (0,0)--(3.5,0);
\draw [ ->, black!70!white] (0,0)--(0,3.5);

\begin{scope}[shift={(2,1)}]

\node[inner sep=0,outer sep=1] (A) at (0,0)  {\footnotesize \blue{$\bullet$}};
\node[inner sep=0,outer sep=1] (B) at (1,0)  {\footnotesize \blue{$\bullet$}};
\node[inner sep=0,outer sep=1] (C) at (1,1) {\footnotesize \blue{$\bullet$}};
\node[inner sep=0,outer sep=1] (D) at (0,1) {\footnotesize \blue{$\bullet$}};

\draw[arrows={-stealth},very thick,blue,below] (A) to node {$\kappa_1$} (B);
\draw[arrows={-stealth},very thick,blue,right] (B) to node {$\kappa_2$} (C);
\draw[arrows={-stealth},very thick,blue,above] (C) to node {$\kappa_3$} (D);
\draw[arrows={-stealth},very thick,blue,left]  (D) to node {$\kappa_4$} (A);

\end{scope}

    \end{tikzpicture}
    \vspace{-0.25cm} 
    \caption{Translating a mass-action system corresponds to multiplying the right-hand side of the dynamical system by a monomial. By translating the system in (a) by $(2,1)^\top$, we obtain the system in (b).
    }
    \label{fig:translation}
\end{figure}

Consider translation of a mass-action system in the plane by a fixed vector $(\alpha, \beta)^\top$. The effect of this operation on the graph corresponds to multiplying the right-hand side of the dynamical system by the monomial $x^\alpha y^\beta$. For example, in \Cref{fig:translation}(a) we have a mass-action system whose dynamical system is
    \eq{ 
        \dot{x} &= \kappa_1-\kappa_3 xy, \\
        \dot{y} &= \kappa_2 x-\kappa_4 y.
    }
We translated this system by $(2,1)^\top$ to obtain the one in \Cref{fig:translation}(b), whose dynamical system is
    \eq{ 
        \dot{x} &= x^2y[\kappa_1-\kappa_3 xy], \\
        \dot{y} &= x^2y[\kappa_2 x-\kappa_4 y].
    }

In higher dimension, translation works similarly. Translating a mass-action system by a fixed vector $(\alpha_1, \alpha_2, \ldots, \alpha_n)^\top$ corresponds to multiplying the right-hand side of the dynamical system by the monomial $x_1^{\alpha_1}x_2^{\alpha_2}\cdots x_n^{\alpha_n}$.
 
\subsection{Scalar multiplication}

\begin{figure}[ht]
    \centering
    \begin{tikzpicture}[scale=0.94]
    \node at (1.75,-0.5) {(a)};

\draw[arrows={-stealth},very thick,white,below] (0,0) to node {\vphantom{$\kappa_1$}} (1,0);

\draw [step=1, gray, very thin] (0,0) grid (3.5,3.5);
\draw [ ->, black!70!white] (0,0)--(3.5,0);
\draw [ ->, black!70!white] (0,0)--(0,3.5);

\begin{scope}[shift={(1,1)}]

\node[inner sep=0,outer sep=1] (A) at (0,0)  {\footnotesize \blue{$\bullet$}};
\node[inner sep=0,outer sep=1] (B) at (1,0)  {\footnotesize \blue{$\bullet$}};
\node[inner sep=0,outer sep=1] (C) at (1,1) {\footnotesize \blue{$\bullet$}};
\node[inner sep=0,outer sep=1] (D) at (0,1) {\footnotesize \blue{$\bullet$}};

\draw[arrows={-stealth},very thick,blue,below] (A) to node {$\kappa_1$} (B);
\draw[arrows={-stealth},very thick,blue,right] (B) to node {$\kappa_2$} (C);
\draw[arrows={-stealth},very thick,blue,above] (C) to node {$\kappa_3$} (D);
\draw[arrows={-stealth},very thick,blue,left]  (D) to node {$\kappa_4$} (A);

\end{scope}

    \end{tikzpicture}
        \hspace{1cm}
    \begin{tikzpicture}[scale=0.94]
    \node at (1.75,-0.5) {(b)};

\draw[arrows={-stealth},very thick,white,below] (0,0) to node {\vphantom{$\kappa_1$}} (1,0);

\draw [step=1, gray, very thin] (0,0) grid (3.5,3.5);
\draw [ ->, black!70!white] (0,0)--(3.5,0);
\draw [ ->, black!70!white] (0,0)--(0,3.5);

\begin{scope}[shift={(1,1)}]

\node[inner sep=0,outer sep=1] (A) at (0,0)  {\footnotesize \blue{$\bullet$}};
\node[inner sep=0,outer sep=1] (B) at (1,0)  {\footnotesize \blue{$\bullet$}};
\node[inner sep=0,outer sep=1] (C) at (1,1) {\footnotesize \blue{$\bullet$}};
\node[inner sep=0,outer sep=1] (D) at (0,1) {\footnotesize \blue{$\bullet$}};

\draw[arrows={-stealth},very thick,blue,below] (A) to node {$5\kappa_1$} (B);
\draw[arrows={-stealth},very thick,blue,right] (B) to node {$5\kappa_2$} (C);
\draw[arrows={-stealth},very thick,blue,above] (C) to node {$5\kappa_3$} (D);
\draw[arrows={-stealth},very thick,blue,left]  (D) to node {$5\kappa_4$} (A);

\end{scope}

    \end{tikzpicture}
        \hspace{1cm}
    \begin{tikzpicture}[scale=0.94]
    \input{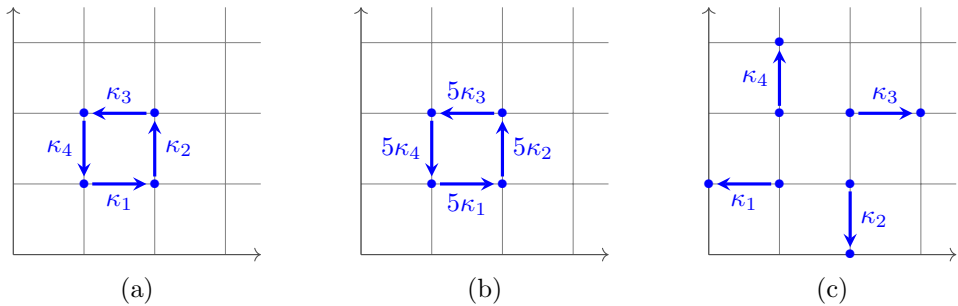}
    \end{tikzpicture}
    \vspace{-0.25cm} 
    \caption{Scalar multiplication of a mass-action system by a positive scalar corresponds to scaling the rate constants, while multiplication by $-1$ corresponds to flipping the reaction vectors. Take the system in (a); multiplying by $5$ we obtain (b), while multiplying by $-1$ we obtain (c).}
    \label{fig:scalar}
\end{figure}

Consider now scalar multiplication of the rate constants by a positive scalar $\lambda$. The effect of this operation on the rate constants corresponds to multiplying the right-hand side of the dynamical system by the constant $\lambda$. For example, in \Cref{fig:scalar}(a) we have a mass-action system whose dynamical system is
    \eq{ 
        \dot{x} &= \kappa_1xy-\kappa_3 x^2y^2, \\ 
        \dot{y} &= \kappa_2 x^2y-\kappa_4 xy^2.
    }
We multiply this system by $5$ to obtain the system in \Cref{fig:scalar}(b), whose dynamical system is
    \eq{ 
        \dot{x} &= 5[\kappa_1xy-\kappa_3 x^2y^2], \\ 
        \dot{y} &= 5[\kappa_2 x^2y-\kappa_4 xy^2].
    }

We can also consider scalar multiplication by the negative scalar $\lambda = -1$, which flips every reaction vectors to the opposite direction, i.e., if a reaction is initially $(\yy, \yy+\vv v)$, then after this operation, we have the reaction $(\yy, \yy-\vv v)$.  
For example, we multiply the system in \Cref{fig:scalar}(a) by $-1$ to obtain the system in \Cref{fig:scalar}(c), whose dynamical system is
    \eq{ 
        \dot{x} &= -[\kappa_1xy-\kappa_3 x^2y^2], \\ 
        \dot{y} &= -[\kappa_2 x^2y-\kappa_4 xy^2].
    }
Note that the graph in \Cref{fig:scalar}(c) does not resemble the original one, and the set of nodes is different.

The multiplication by a negative scalar $\lambda$ is seen as first flipping the reaction vectors, then scale the rate constants by $|\lambda|$.

In higher dimension, scalar multiplication works the same way.

\subsection{Addition}

\begin{figure}[ht]
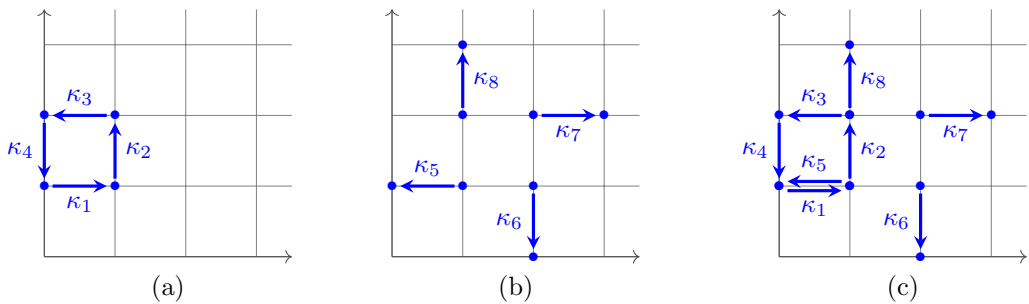

    \centering
    \begin{tikzpicture}[scale=0.94]
    \node at (1.75,-0.45) {(a)};

\draw[arrows={-stealth},very thick,white,below] (0,0) to node {\vphantom{$\kappa_1$}} (1,0);

\draw [step=1, gray, very thin] (0,0) grid (3.5,3.5);
\draw [ ->, black!70!white] (0,0)--(3.5,0);
\draw [ ->, black!70!white] (0,0)--(0,3.5);

\begin{scope}[shift={(0,1)}]

\node[inner sep=0,outer sep=1] (A) at (0,0)  {\footnotesize \blue{$\bullet$}};
\node[inner sep=0,outer sep=1] (B) at (1,0)  {\footnotesize \blue{$\bullet$}};
\node[inner sep=0,outer sep=1] (C) at (1,1) {\footnotesize \blue{$\bullet$}};
\node[inner sep=0,outer sep=1] (D) at (0,1) {\footnotesize \blue{$\bullet$}};

\draw[arrows={-stealth},very thick,blue,below] (A) to node {$\kappa_1$} (B);
\draw[arrows={-stealth},very thick,blue,right] (B) to node {$\kappa_2$} (C);
\draw[arrows={-stealth},very thick,blue,above] (C) to node {$\kappa_3$} (D);
\draw[arrows={-stealth},very thick,blue,left]  (D) to node {$\kappa_4$} (A);
\end{scope}

    \end{tikzpicture}
        \hspace{1cm}
    \begin{tikzpicture}[scale=0.94]
    \input{fig-unitsq-add2.tex}
    \end{tikzpicture}
        \hspace{1cm}
    \begin{tikzpicture}[scale=0.94]
    \input{fig-unitsq-add3.tex}
    \end{tikzpicture}
    \vspace{-0.25cm} 
    \caption{Addition of mass-action systems corresponds to adding together the right-hand sides of the dynamical systems. By adding the systems in (a) and (b), we obtain the system in (c).}
    \label{fig:addition}
\end{figure}

Consider the addition of mass-action systems, where we take the union of the sets of vertices and the sets of edges, respectively. The rate constants are treated additively, as shown in the example below. The effect of this operation corresponds to adding together the right-hand sides of the two dynamical systems. For example, in \Cref{fig:addition}(a) and \Cref{fig:addition}(b), we have two mass-action systems whose associated dynamical systems are
    \eq{ 
        \begin{array}{l}
            \dot{x} = \kappa_1y-\kappa_3 xy^2 \\
            \dot{y} = \kappa_2 xy-\kappa_4 y^2
        \end{array}
        \quad \text{and} \quad 
        \begin{array}{l}
           \dot{x} = -\kappa_5xy + \kappa_7 x^2y^2 \\ 
            \dot{y} = -\kappa_6 x^2y + \kappa_8 xy^2
        \end{array}
        .
    }
We add these systems to obtain the one in \Cref{fig:addition}(c), whose dynamical system is
    \eq{ 
        \dot{x} &= [\kappa_1y-\kappa_3 xy^2] + [-\kappa_5xy + \kappa_7 x^2y^2],\\
        \dot{y} &= [\kappa_2 xy-\kappa_4 y^2] + [-\kappa_6 x^2y + \kappa_8 xy^2].
    }

In higher dimension, addition works exactly the same.

\section{Examples}
\label{sec:examples}

We now apply a combination of the above operations to create a general procedure that gives rise to weakly reversible mass-action systems with infinitely many positive steady states. We start by picking a weakly reversible mass-action system, which we call the \df{base unit}. Take a negative scalar multiple of the base unit. Each reaction in this (necessarily not weakly reversible) system came from a reaction in the base unit by reflection. Therefore, as in \Cref{fig:addition}, we can add an appropriate translated copy of the base unit, so the two reactions form a reversible pair. By adding the negative scalar multiple and all the translated copies, we arrive at a weakly reversible mass-action system. Its associated dynamical system is obtained from that of the base unit by multiplying each equation by the same scalar polynomial. With the monomials in the common factor represented by their exponent vectors as points in $\rr^n$, by construction the negative term is in the interior of the convex hull of the positive terms, an object called the Newton polytope \cite{sturmfels:1996}. Therefore, the value of the polynomial near the boundary of $\rr_+^n$ is positive (see \cite{pantea:koeppl:craciun:2012} for a detailed study of monomial domination). If the coefficient of the negative term is sufficiently negative, the value of the polynomial at $(1,1,\ldots,1)^\top$ is negative. Consequently, this polynomial has uncountably many roots in $\rr_+^n$, resulting in infinitely many positive steady states of the constructed mass-action system.

\begin{ex}
\label{ex:main}
    The square in \Cref{fig:translation}(a) with rate constants $\kk_i = 1$ is our base unit, and \Cref{fig:ex-main-pieces} shows five transformed copies of it. The full system obtained by adding these five together (\Cref{fig:ex-main-complete}) is a weakly reversible mass-action system whose associated dynamical system is
    \eq{
        \begin{split}
        \dot{x} &= (x^2y + xy^2 + x + y - 5xy)[1-xy], \\
        \dot{y} &= (x^2y + xy^2 + x + y - 5xy)[x-y].
        \end{split}
    }
    The set of positive steady states is $\{ (x,y)^\top\in\rrpp^2\colon x^2y + xy^2 + x + y-5xy = 0 \} \cup \{(1,1)^\top\}$. The first piece is a closed curve in $\rrpp^2$, and the point $(1,1)^\top$ is the unique positive steady state of the base unit. The phase portraits of the base unit and the full system are shown in \Cref{fig:streamplots_cycle}.
    
    An analogous example exists in any dimension $n\geq2$, resulting in a hypersurface of positive steady states. There the base unit is a weakly reversible $n$-cube, and the common factor has $2n$ terms with coefficient $1$ and the monomial $-\alpha x_1x_2\cdots x_n$ with $\alpha>2n$. For example, in the $n=3$ case the common factor is $x^2yz + xy^2z + xyz^2 + yz + xz + xy - \alpha xyz$ with $\alpha>6$.

    \begin{figure}[htbp]
        \centering
        \begin{tikzpicture}[scale=0.94]
        \input{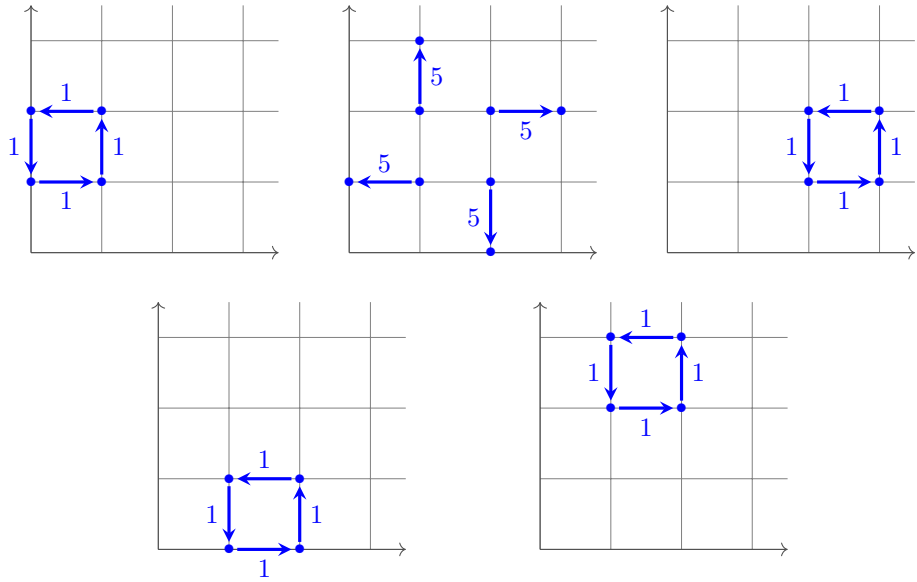}
        \end{tikzpicture}
        \caption{The five building blocks of \Cref{ex:main}, a weakly reversible mass-action system with infinitely many positive steady states as shown in \Cref{fig:ex-main-complete}.}
        \label{fig:ex-main-pieces}
    \end{figure}
    
    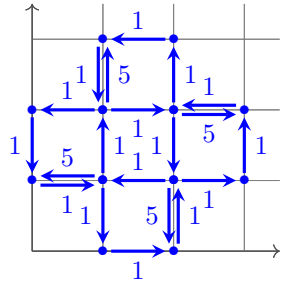
\begin{figure}[htbp]
        \centering
        \begin{tikzpicture}[scale=0.94]
        \draw [step=1, gray, very thin] (0,0) grid (3.5,3.5);
\draw [ ->, black!70!white] (0,0)--(3.5,0);
\draw [ ->, black!70!white] (0,0)--(0,3.5);

\draw [step=1, gray, very thin] (0,0) grid (3.5,3.5);
\draw [ ->, black!70!white] (0,0)--(3.5,0);
\draw [ ->, black!70!white] (0,0)--(0,3.5);

\begin{scope}[shift={(0,1)}]
\node[inner sep=0,outer sep=1] (A) at (0,0)  {\footnotesize \blue{$\bullet$}};
\node[inner sep=0,outer sep=1] (B) at (1,0)  {\footnotesize \blue{$\bullet$}};
\node[inner sep=0,outer sep=1] (C) at (1,1) {\footnotesize \blue{$\bullet$}};
\node[inner sep=0,outer sep=1] (D) at (0,1) {\footnotesize \blue{$\bullet$}};

\draw[arrows={-stealth},very thick,blue,below, transform canvas={yshift=-1.75pt}] (A) to node {$1$} (B);
\draw[arrows={-stealth},very thick,blue,right] (B) to node {$1$} (C);
\draw[arrows={-stealth},very thick,blue,above] (C) to node {$1$} (D);
\draw[arrows={-stealth},very thick,blue,left]  (D) to node {$1$} (A);

\draw[arrows={-stealth},very thick,blue,above, transform canvas={yshift=+1.75pt}] (B) to node {$5$} (A);
\end{scope}

\begin{scope}[shift={(2,1)}]
\node[inner sep=0,outer sep=1] (A) at (0,0)  {\footnotesize \blue{$\bullet$}};
\node[inner sep=0,outer sep=1] (B) at (1,0)  {\footnotesize \blue{$\bullet$}};
\node[inner sep=0,outer sep=1] (C) at (1,1) {\footnotesize \blue{$\bullet$}};
\node[inner sep=0,outer sep=1] (D) at (0,1) {\footnotesize \blue{$\bullet$}};

\draw[arrows={-stealth},very thick,blue,below] (A) to node {$1$} (B);
\draw[arrows={-stealth},very thick,blue,right] (B) to node {$1$} (C);
\draw[arrows={-stealth},very thick,blue,above, transform canvas={yshift=+1.75pt}] (C) to node {$1$} (D);
\draw[arrows={-stealth},very thick,blue,left]  (D) to node {$1$} (A);

\draw[arrows={-stealth},very thick,blue,below, transform canvas={yshift=-1.75pt}] (D) to node {$5$} (C);
\end{scope}


\begin{scope}[shift={(1,0)}]
\node[inner sep=0,outer sep=1] (A) at (0,0)  {\footnotesize \blue{$\bullet$}};
\node[inner sep=0,outer sep=1] (B) at (1,0)  {\footnotesize \blue{$\bullet$}};
\node[inner sep=0,outer sep=1] (C) at (1,1) {\footnotesize \blue{$\bullet$}};
\node[inner sep=0,outer sep=1] (D) at (0,1) {\footnotesize \blue{$\bullet$}};

\draw[arrows={-stealth},very thick,blue,below] (A) to node {$1$} (B);
\draw[arrows={-stealth},very thick,blue,right,transform canvas={xshift=1.75pt}] (B) to node {$1$} (C);
\draw[arrows={-stealth},very thick,blue,above] (C) to node {$1$} (D);
\draw[arrows={-stealth},very thick,blue,left]  (D) to node {$1$} (A);

\draw[arrows={-stealth},very thick,blue,left, transform canvas={xshift=-1.75pt}] (C) to node {$5$} (B);
\end{scope}

\begin{scope}[shift={(1,2)}]
\node[inner sep=0,outer sep=1] (A) at (0,0)  {\footnotesize \blue{$\bullet$}};
\node[inner sep=0,outer sep=1] (B) at (1,0)  {\footnotesize \blue{$\bullet$}};
\node[inner sep=0,outer sep=1] (C) at (1,1) {\footnotesize \blue{$\bullet$}};
\node[inner sep=0,outer sep=1] (D) at (0,1) {\footnotesize \blue{$\bullet$}};

\draw[arrows={-stealth},very thick,blue,below] (A) to node {$1$} (B);
\draw[arrows={-stealth},very thick,blue,right] (B) to node {$1$} (C);
\draw[arrows={-stealth},very thick,blue,above] (C) to node {$1$} (D);
\draw[arrows={-stealth},very thick,blue,left,transform canvas={xshift=-1.75pt}]  (D) to node {$1$} (A);

\draw[arrows={-stealth},very thick,blue,right, transform canvas={xshift=+1.75pt}] (A) to node {$5$} (D);
\end{scope}
        \end{tikzpicture}
        \caption{The full system of \Cref{ex:main}, a weakly reversible mass-action system with infinitely many positive steady states.}
        \label{fig:ex-main-complete}
    \end{figure}

    \begin{figure}[htbp]
    \centering
    \vspace{-0.5cm} 
    \begin{tikzpicture}
        \node at (0,0) [above=-8.5pt] {\includegraphics[width=0.39\textwidth]{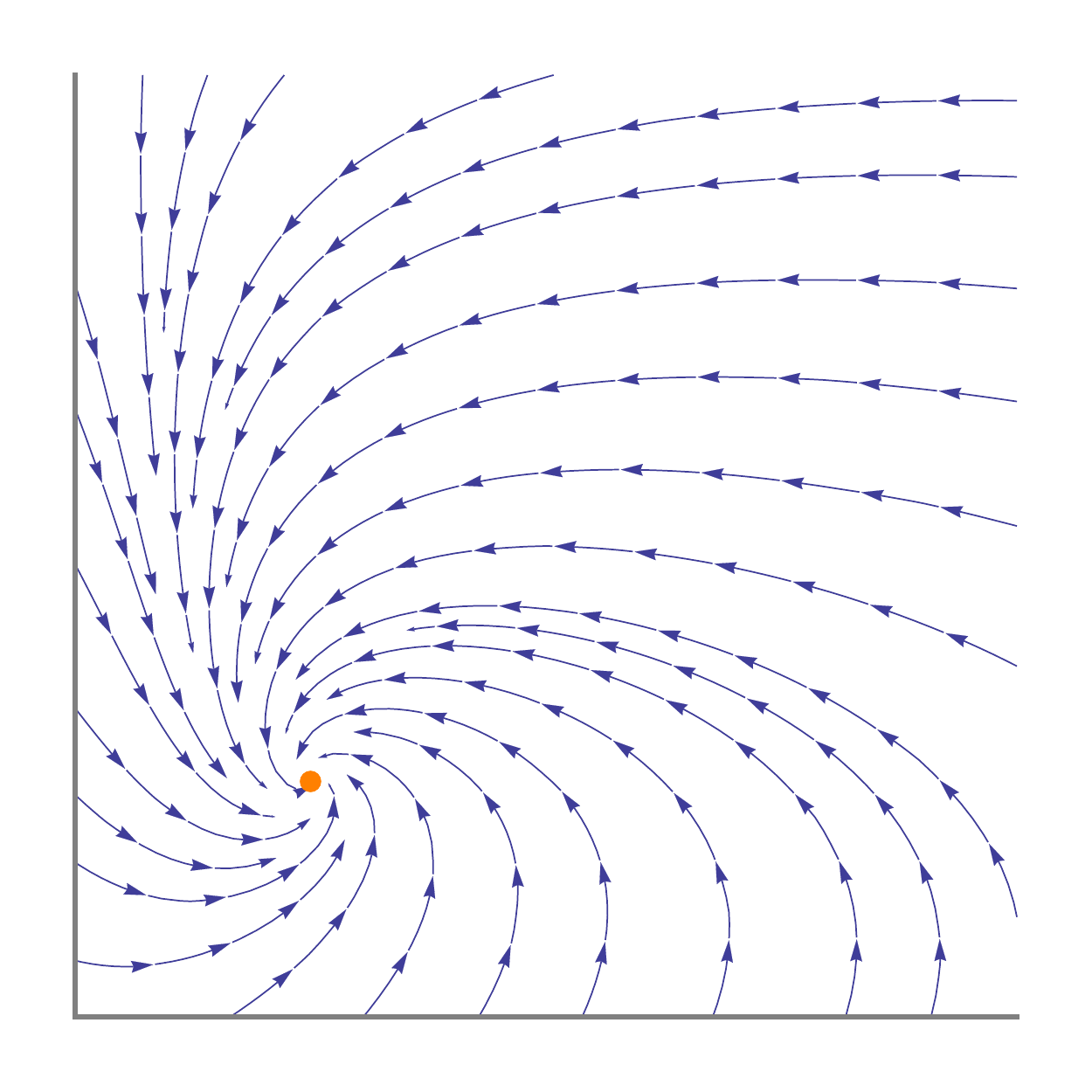} };
        \node at (0,-0.45) {(a)};
    \end{tikzpicture}
    \hspace{0.25cm}
    \begin{tikzpicture}
        \node at (0,0) [above=-8.5pt] {\includegraphics[width=0.39\textwidth]{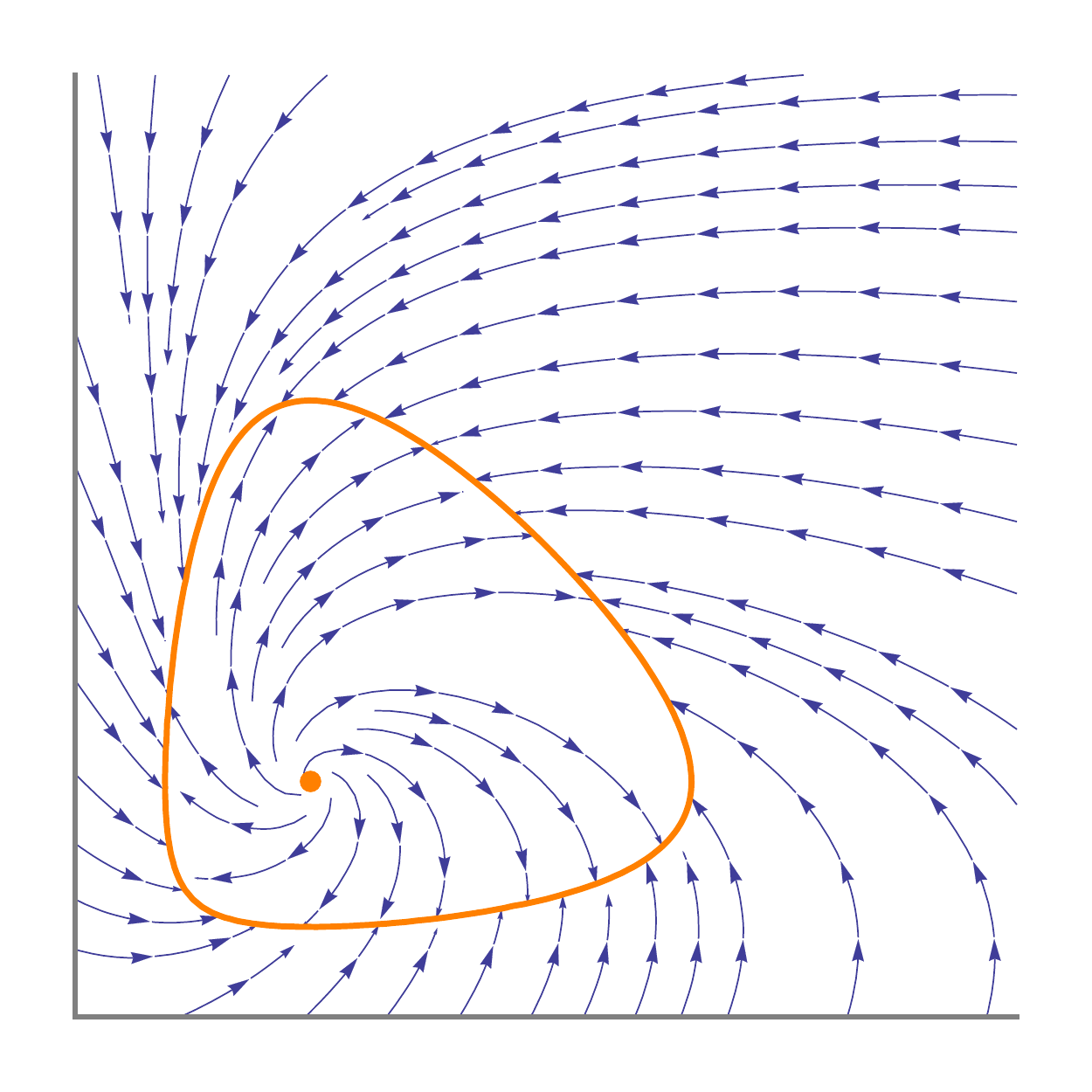}};
        \node at (0,-0.45) {(b)};
    \end{tikzpicture}
    \vspace{-0.25cm} 
    \caption{The phase portraits of the base unit (a) and the full system (b), from \Cref{ex:main}. The full system has a curve of positive steady states, shown in orange.}
        \label{fig:streamplots_cycle}
    \end{figure}
\end{ex}

\newpage
\begin{ex}
\label{ex:cross}
    Consider the reversible mass-action system in \Cref{fig:cross}(a). We saw this example in \Cref{sec:intro}. Its associated dynamical system is 
    \eq{ 
        \dot{x} &= (x^2y^2+x^2+y^2+1-5xy)[1-xy+y-x], \\
        \dot{y} &= (x^2y^2+x^2+y^2+1-5xy)[1-xy-y+x].
    }
    The common factor $x^2y^2+x^2+y^2+1-5xy$ results in a curve of positive steady states, depicted in orange in \Cref{fig:cross}(b).
    
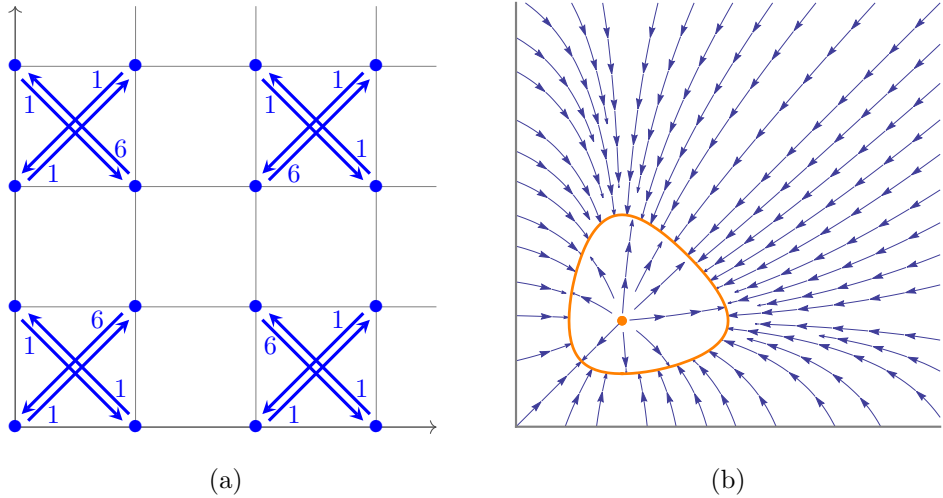
\begin{figure}[htb!]
    \centering
    \vspace{-0.5cm} 
    \begin{tikzpicture}[scale=1.6]
    \node at (1.75,-0.45) {(a)};

\draw [white] (0,0)--(1,0) node [below] {\vphantom{$1$}};
\node at (0,-0.2) {\vphantom{.}};

\draw [step=1, gray, very thin] (0,0) grid (3.5,3.5);
\draw [ ->, black!70!white] (0,0)--(3.5,0);
\draw [ ->, black!70!white] (0,0)--(0,3.5);

\begin{scope}[shift={(0,0)}]
\node[inner sep=0,outer sep=1] (A) at (0,0)  {\large \blue{$\bullet$}};
\node[inner sep=0,outer sep=1] (B) at (1,0) {\large \blue{$\bullet$}};
\node[inner sep=0,outer sep=1] (C) at (1,1) {\large \blue{$\bullet$}};
\node[inner sep=0,outer sep=1] (D) at (0,1) {\large \blue{$\bullet$}};

\draw[arrows={-stealth},very thick,blue, transform canvas={xshift=1.24pt, yshift=1.24pt}] (B) to node [near start, right] {$1$} (D);
\draw[arrows={-stealth},very thick,blue , transform canvas={xshift=-1.24pt, yshift=-1.24pt}] (D) to node [near start,left] {$1$} (B);

\draw[arrows={-stealth},very thick,blue, transform canvas={xshift=-1.24pt, yshift=1.24pt}] (C) to node [near start,above] {$6$} (A);
\draw[arrows={-stealth},very thick,blue , transform canvas={xshift=1.24pt, yshift=-1.24pt}] (A) to node[near start,below] {$1$} (C);

\end{scope}

\begin{scope}[shift={(2,0)}]
\node[inner sep=0,outer sep=1] (A) at (0,0)  {\large \blue{$\bullet$}};
\node[inner sep=0,outer sep=1] (B) at (1,0) {\large \blue{$\bullet$}};
\node[inner sep=0,outer sep=1] (C) at (1,1) {\large \blue{$\bullet$}};
\node[inner sep=0,outer sep=1] (D) at (0,1) {\large \blue{$\bullet$}};

\draw[arrows={-stealth},very thick,blue, transform canvas={xshift=1.24pt, yshift=1.24pt}] (B) to node [near start, right] {$1$} (D);
\draw[arrows={-stealth},very thick,blue , transform canvas={xshift=-1.24pt, yshift=-1.24pt}] (D) to node [near start,left] {$6$} (B);

\draw[arrows={-stealth},very thick,blue, transform canvas={xshift=-1.24pt, yshift=1.24pt}] (C) to node [near start,above] {$1$} (A);
\draw[arrows={-stealth},very thick,blue , transform canvas={xshift=1.24pt, yshift=-1.24pt}] (A) to node[near start,below] {$1$} (C);

\end{scope}

\begin{scope}[shift={(2,2)}]
\node[inner sep=0,outer sep=1] (A) at (0,0)  {\large \blue{$\bullet$}};
\node[inner sep=0,outer sep=1] (B) at (1,0) {\large \blue{$\bullet$}};
\node[inner sep=0,outer sep=1] (C) at (1,1) {\large \blue{$\bullet$}};
\node[inner sep=0,outer sep=1] (D) at (0,1) {\large \blue{$\bullet$}};

\draw[arrows={-stealth},very thick,blue, transform canvas={xshift=1.24pt, yshift=1.24pt}] (B) to node [near start, right] {$1$} (D);
\draw[arrows={-stealth},very thick,blue , transform canvas={xshift=-1.24pt, yshift=-1.24pt}] (D) to node [near start,left] {$1$} (B);

\draw[arrows={-stealth},very thick,blue, transform canvas={xshift=-1.24pt, yshift=1.24pt}] (C) to node [near start,above] {$1$} (A);
\draw[arrows={-stealth},very thick,blue , transform canvas={xshift=1.24pt, yshift=-1.24pt}] (A) to node[near start,below] {$6$} (C);

\end{scope}

\begin{scope}[shift={(0,2)}]
\node[inner sep=0,outer sep=1] (A) at (0,0)  {\large \blue{$\bullet$}};
\node[inner sep=0,outer sep=1] (B) at (1,0) {\large \blue{$\bullet$}};
\node[inner sep=0,outer sep=1] (C) at (1,1) {\large \blue{$\bullet$}};
\node[inner sep=0,outer sep=1] (D) at (0,1) {\large \blue{$\bullet$}};

\draw[arrows={-stealth},very thick,blue, transform canvas={xshift=1.24pt, yshift=1.24pt}] (B) to node [near start, right] {$6$} (D);
\draw[arrows={-stealth},very thick,blue , transform canvas={xshift=-1.24pt, yshift=-1.24pt}] (D) to node [near start,left] {$1$} (B);

\draw[arrows={-stealth},very thick,blue, transform canvas={xshift=-1.24pt, yshift=1.24pt}] (C) to node [near start,above] {$1$} (A);
\draw[arrows={-stealth},very thick,blue , transform canvas={xshift=1.24pt, yshift=-1.24pt}] (A) to node[near start,below] {$1$} (C);

\end{scope}

    \end{tikzpicture}
    \hspace{0.25cm}
    \begin{tikzpicture}
        \node at (0,0) [above=-8.5pt] {\includegraphics[width=0.39\textwidth]{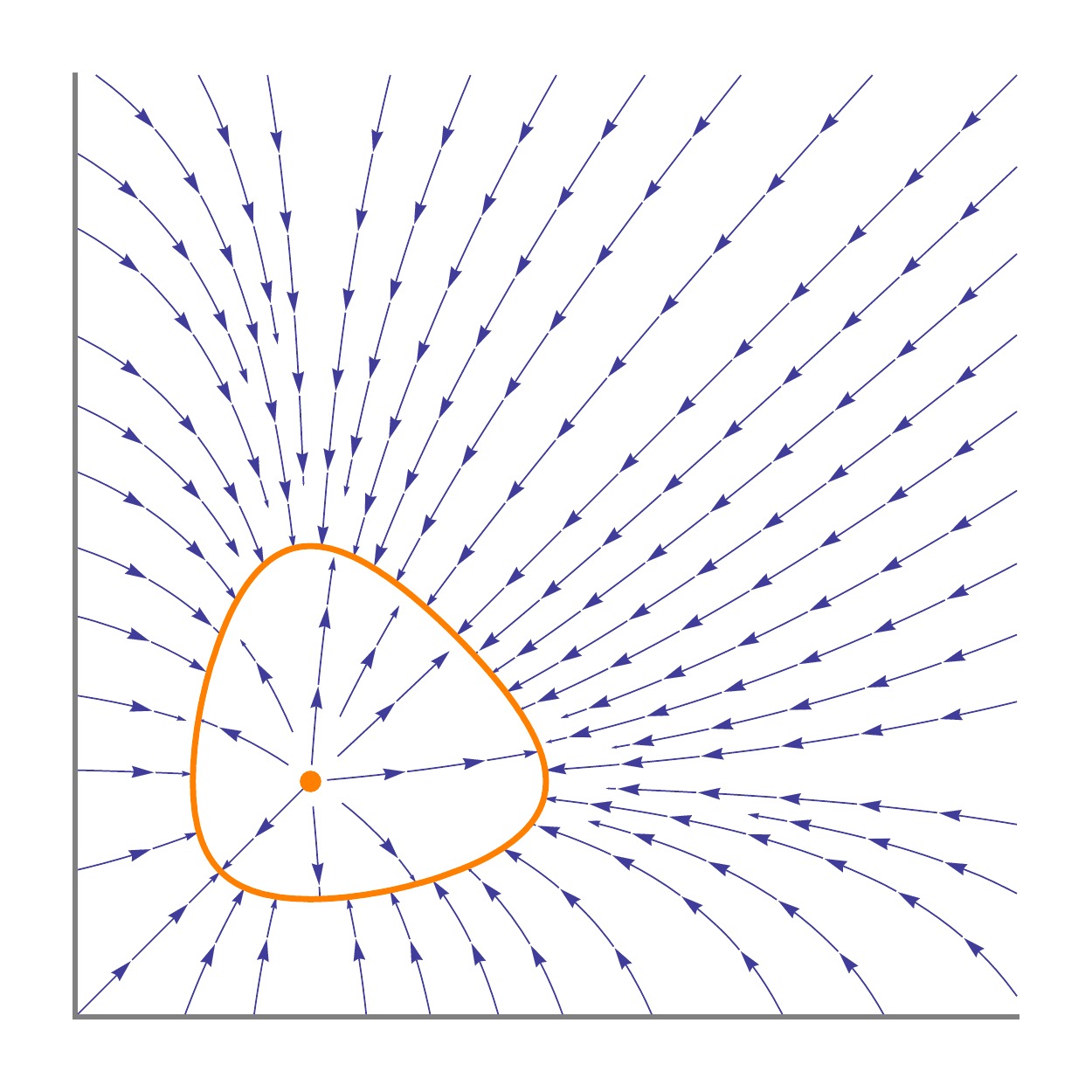}};
        \node at (0,-0.45) {(b)};
    \end{tikzpicture}
    \vspace{-0.25cm}    
    \caption{The mass-action system of \Cref{ex:cross} and its phase portrait. Shown in orange is the set of positive steady states.}
    \label{fig:cross}
\end{figure}
\end{ex}


    
    
    


\begin{ex}
\label{ex:triangle}
    Consider the weakly reversible mass-action system as depicted in \Cref{fig:triangle}(a). Its associated dynamical system is 
    \eq{ 
    \dot{x} &= (x^2+xy^2+y-4xy)[1-x], \\
    \dot{y} &= (x^2+xy^2+y-4xy)[x-y].
    }
    The common factor $x^2+xy^2+y-4xy$ results in a curve of positive steady states, depicted in orange in \Cref{fig:triangle}(b). Note that the polynomial on the right-hand side has degree $4$.

\begin{figure}[htb!]
    \centering
    \vspace{-0.5cm} 
    \begin{tikzpicture}[scale=1.6]
    \node at (1.75,-0.45) {(a)};

\node at (0,-0.2) {\vphantom{.}};

\draw [step=1, gray, very thin] (0,0) grid (3.5,3.5);
\draw [ ->, black!70!white] (0,0)--(3.5,0);
\draw [ ->, black!70!white] (0,0)--(0,3.5);

\begin{scope}[shift={(0,1)}]
\node[inner sep=0,outer sep=1] (A) at (0,0)  {\large \blue{$\bullet$}};
\node[inner sep=0,outer sep=1] (B) at (1,0) {\large \blue{$\bullet$}};
\node[inner sep=0,outer sep=1] (C) at (0,1) {\large \blue{$\bullet$}};

\draw[arrows={-stealth},very thick,blue,below, transform canvas={yshift=-1.75pt}] (B) to node {$4$} (A);
    \draw[arrows={-stealth},very thick,blue,above, transform canvas={yshift=1.75pt}] (A) to node {$1$} (B);
\draw[arrows={-stealth},very thick,blue,above right] (B) to node {\!\!$1$} (C);
\draw[arrows={-stealth},very thick,blue,left] (C) to node {$1$} (A);
\end{scope}

\begin{scope}[shift={(1,2)}]
\node[inner sep=0,outer sep=1] (A) at (0,0)  {\large \blue{$\bullet$}};
\node[inner sep=0,outer sep=1] (B) at (1,0) {\large \blue{$\bullet$}};
\node[inner sep=0,outer sep=1] (C) at (0,1) {\large \blue{$\bullet$}};

\draw[arrows={-stealth},very thick,blue,below] (A) to node {$1$} (B);
\draw[arrows={-stealth},very thick,blue,above right] (B) to node {\!\!$1$} (C);
\draw[arrows={-stealth},very thick,blue,left, transform canvas={xshift=-1.75pt}] (A) to node {$4$} (C);
    \draw[arrows={-stealth},very thick,blue,right, transform canvas={xshift=1.75pt}] (C) to node {$1$} (A);
\end{scope}

\begin{scope}[shift={(2,0)}]
\node[inner sep=0,outer sep=1] (A) at (0,0)  {\large \blue{$\bullet$}};
\node[inner sep=0,outer sep=1] (B) at (1,0) {\large \blue{$\bullet$}};
\node[inner sep=0,outer sep=1] (C) at (0,1) {\large \blue{$\bullet$}};

\draw[arrows={-stealth},very thick,blue,below] (A) to node {$1$} (B);
\draw[arrows={-stealth},very thick,blue,above right, transform canvas={xshift=1.24pt, yshift=1.24pt}] (C) to node {\!\!$4$} (B);
    \draw[arrows={-stealth},very thick,blue,below left, transform canvas={xshift=-1.24pt, yshift=-1.24pt}] (B) to node {$1$\!\!} (C);
\draw[arrows={-stealth},very thick,blue,left] (C) to node {$1$} (A);
\end{scope}

    \end{tikzpicture}
    \hspace{0.25cm}
    \begin{tikzpicture}
        \node at (0,0) [above=-8.5pt] {\includegraphics[width=0.39\textwidth]{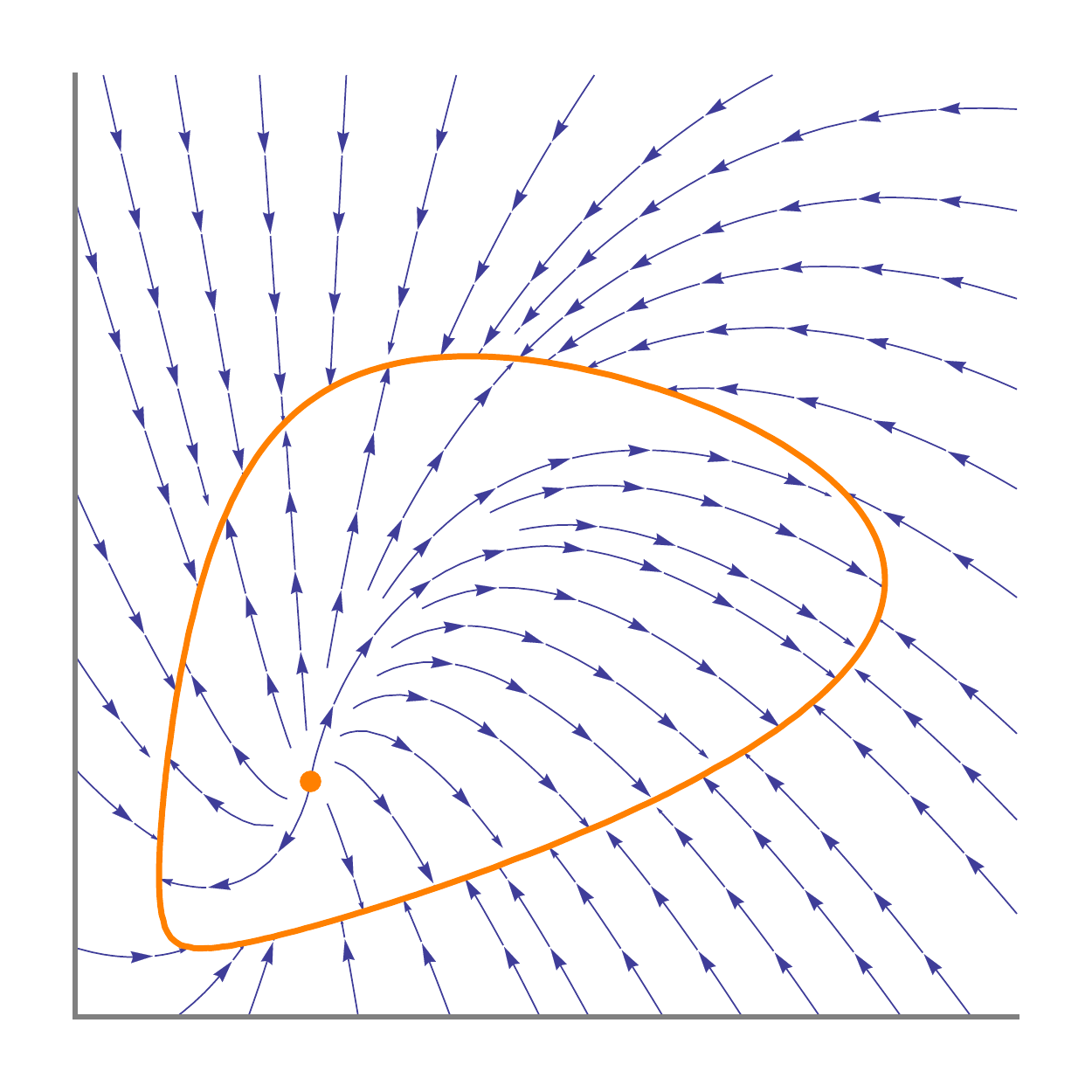}};
        \node at (0,-0.45) {(b)};
    \end{tikzpicture}
    \vspace{-0.25cm}
    \caption{The mass-action system of \Cref{ex:triangle} and its phase portrait. Shown in orange is the set of positive steady states.}
    \label{fig:triangle}
\end{figure}
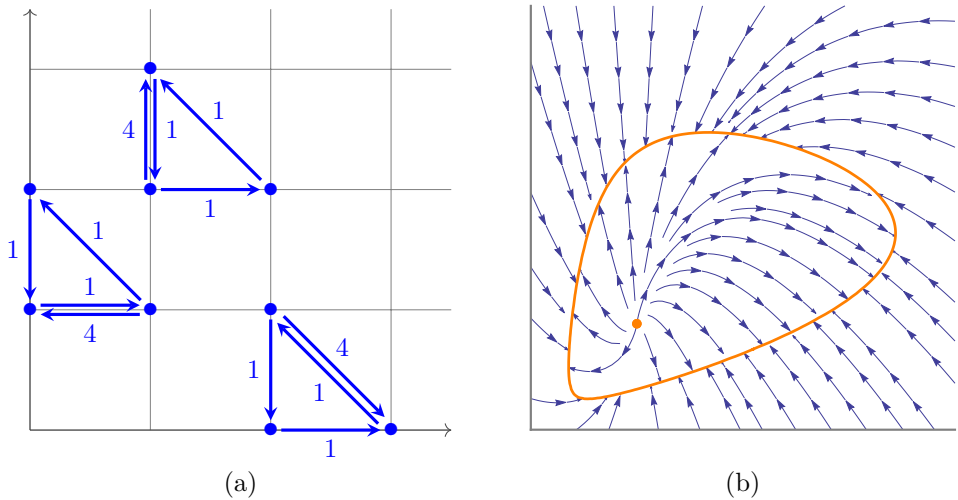

\end{ex}

\section{Open questions and discussion}
\label{sec:open}

In this section we list a couple of open questions. Below, we are asking the questions in the context of constructing a (weakly) reversible mass-action system with infinitely many positive steady states.  

\begin{itemize}
    \item What is the minimal degree of the polynomials in the right-hand side of the dynamical system?
    \item What is the minimum number of vertices?
    \item What is the minimum number of reactions?
    \item What is the minimum deficiency?
\end{itemize}

The \df{deficiency} of a reaction network is the nonnegative integer $m-\ell-\dim S$, where $m$ is the number of vertices, $\ell$ is the number of connected components, and $S$ is the stoichiometric subspace \cite{feinberg:1972}. Pertaining to the above questions, note that \Cref{ex:triangle} has as the right-hand side of its dynamical system, a degree-$4$ polynomial system. It has $9$ vertices, $12$ reactions, and its deficiency is $4$.

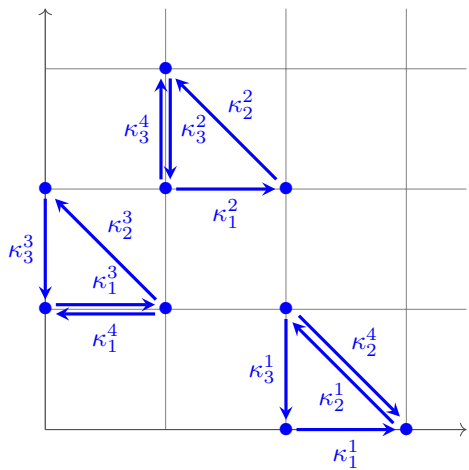
\begin{figure}[htb!]
    \centering
    \begin{tikzpicture}[scale=1.6]
    \node at (0,-0.2) {\vphantom{.}};

\draw [step=1, gray, very thin] (0,0) grid (3.5,3.5);
\draw [ ->, black!70!white] (0,0)--(3.5,0);
\draw [ ->, black!70!white] (0,0)--(0,3.5);

\begin{scope}[shift={(0,1)}]
\node[inner sep=0,outer sep=1] (A) at (0,0)  {\large \blue{$\bullet$}};
\node[inner sep=0,outer sep=1] (B) at (1,0) {\large \blue{$\bullet$}};
\node[inner sep=0,outer sep=1] (C) at (0,1) {\large \blue{$\bullet$}};

\draw[arrows={-stealth},very thick,blue,below, transform canvas={yshift=-1.75pt}] (B) to node {$\kk^{4}_1$} (A);
    \draw[arrows={-stealth},very thick,blue,above, transform canvas={yshift=1.75pt}] (A) to node {$\kk^{3}_1$} (B);
\draw[arrows={-stealth},very thick,blue,above right] (B) to node {\!\!$\kk^{3}_2$} (C);
\draw[arrows={-stealth},very thick,blue,left] (C) to node {$\kk^{3}_3$} (A);
\end{scope}

\begin{scope}[shift={(1,2)}]
\node[inner sep=0,outer sep=1] (A) at (0,0)  {\large \blue{$\bullet$}};
\node[inner sep=0,outer sep=1] (B) at (1,0) {\large \blue{$\bullet$}};
\node[inner sep=0,outer sep=1] (C) at (0,1) {\large \blue{$\bullet$}};

\draw[arrows={-stealth},very thick,blue,below] (A) to node {$\kk^{2}_1$} (B);
\draw[arrows={-stealth},very thick,blue,above right] (B) to node {\!\!$\kk^{2}_2$} (C);
\draw[arrows={-stealth},very thick,blue,left, transform canvas={xshift=-1.75pt}] (A) to node {$\kk^{4}_3$} (C);
    \draw[arrows={-stealth},very thick,blue,right, transform canvas={xshift=1.75pt}] (C) to node {$\kk^{2}_3$} (A);
\end{scope}

\begin{scope}[shift={(2,0)}]
\node[inner sep=0,outer sep=1] (A) at (0,0)  {\large \blue{$\bullet$}};
\node[inner sep=0,outer sep=1] (B) at (1,0) {\large \blue{$\bullet$}};
\node[inner sep=0,outer sep=1] (C) at (0,1) {\large \blue{$\bullet$}};

\draw[arrows={-stealth},very thick,blue,below] (A) to node {$\kk^{1}_1$} (B);
\draw[arrows={-stealth},very thick,blue,above right, transform canvas={xshift=1.24pt, yshift=1.24pt}] (C) to node {\!\!$\kk^{4}_2$} (B);
    \draw[arrows={-stealth},very thick,blue,below left, transform canvas={xshift=-1.24pt, yshift=-1.24pt}] (B) to node {$\kk^{1}_2$\!\!\!} (C);
\draw[arrows={-stealth},very thick,blue,left] (C) to node {$\kk^{1}_3$} (A);
\end{scope}

    \end{tikzpicture}
    \caption{The network of \Cref{ex:triangle} with more general rate constants.}
    \label{fig:triangle-general}
\end{figure}

The rate constants in the examples of \Cref{sec:examples} admit some flexibility, though they cannot be set arbitrarily. On one hand, we can always choose rate constants for a weakly reversible network such that the resulting mass-action system is complex-balanced. In particular, it has a unique positive steady state in each invariant polyhedron. On the other hand, for the system in \Cref{fig:triangle-general}, if 
\eq{
\frac{\kappa_2^1}{\kappa_1^1}=\frac{\kappa_2^2}{\kappa_1^2}=\frac{\kappa_2^3}{\kappa_1^3}=\frac{\kappa_2^4}{\kappa_1^4} \text{\quad and \quad} 
\frac{\kappa_3^1}{\kappa_2^1}=\frac{\kappa_3^2}{\kappa_2^2}=\frac{\kappa_3^3}{\kappa_2^3}=\frac{\kappa_3^4}{\kappa_2^4},
}
then there is a common factor $\kappa_1^1 x^2 + \kappa_1^2 xy^2 + \kappa_1^3 y - \kappa_1^4 xy$, resulting in a continuum of positive steady states provided $\kappa_1^4$ is sufficiently large. In light of this, we pose the following question:
\begin{itemize}
    \item Is it possible for a weakly reversible network to have infinitely many positive steady states for each choice of rate constants in an open set of the parameter space $\mathbb{R}^E_+$?
\end{itemize}

Unlike all of our examples, a common factor is in general not necessary for a continuum of positive roots in 3 dimensions or higher. So the interesting question is:
\begin{itemize}
    \item Can a weakly reversible mass-action system have infinitely many positive steady states without having a common factor on the right-hand side of its differential equations?
\end{itemize}

The analysis and characterization of the set of steady states is one of the most fundamental questions about mathematical models of reaction or interaction networks. In particular, in order to best interpret data from experimental observations, and in order to design software that finds the steady states of a specific model, it is important to know if we should expect the set of fixed points to be finite, or discrete, or a continuum \cites{gunawardena:2003,karp:perezmillan:dasgupta:dickenstein:gunawardena:2012}. 

Mass-action models of reversible and weakly reversible reaction networks are known or conjectured to enjoy several remarkable properties (such as existence of positive equilibria for any parameter values \cite{boros:2019}, persistence \cites{craciun:nazarov:pantea:2013,pantea:2012}, and permanence \cites{boros:hofbauer:2019,gopalkrishnan:miller:shiu:2014,brunner:craciun:2019}). It has also been widely believed that they can have at most a \emph{finite} number of positive steady states within any invariant polyhedron, but here we have shown that this is actually \emph{not true}. Hence, it becomes important to identify minimal additional assumptions (e.g. low degree, sparsity, or low deficiency of the network) that, in addition to reversibility, do actually imply that the set of positive steady states is finite. While low degree may be a good candidate for such an assumption, note that high degree models do appear naturally in applications, for example via elimination of denominators for Michaelis-Menten or Hill kinetics \cite{brunner:craciun:2019}; therefore, a better option may be some version of sparsity or low deficiency \cites{gunawardena:2003,craciun:yu:2018}.

\section*{Acknowledgements} 

The authors started this work during the 2nd Madison Workshop on Mathematics of Reaction Networks in 2018. BB was partially supported by the Austrian Science Fund (FWF), project P28406. GC and PYY were supported in part by the National Science Foundation (NSF) under grant DMS-1816238. PYY was supported in part by the Natural Sciences and Engineering Research Council of Canada (NSERC) PGS-D. We appreciate the various useful suggestions by the anonymous referees.

\bibliographystyle{amsxport}
\bibliography{cit}

\end{document}